\documentclass[preprint]{aastex}

\newcommand{\etal}{et al.\ }
\newcommand{\kms}{\, {\rm km\, s}^{-1}}
\newcommand{\mpc}{\, {\rm Mpc}}
\newcommand{\kpc}{\, {\rm Kpc}}
\newcommand{\hmpc}{\, h^{-1} \mpc}
\newcommand{\hkpc}{\, h^{-1} \kpc}
\newcommand{\lya}{Ly$\alpha$ }

\newcommand{\gmo}{\gamma-1}
\newcommand{\bF}{\bar{F}}

\slugcomment{Submitted to ApJL}

\shorttitle{Evolution of the Ionizing Background}

\shortauthors{McDonald \etal}

\begin{document}

\title{The \lya Forest Flux Distribution at $z\sim 5.2$ and
the Evolution of the Ionizing Background}

\author{Patrick McDonald\altaffilmark{1,2} and
Jordi Miralda-Escud\'e \altaffilmark{1,3}
}
\altaffiltext{1}{Department of Astronomy, The Ohio State University, 
Columbus, OH 43210; mcdonald,jordi@astronomy.ohio-state.edu}
\altaffiltext{2}{Department of Physics and Astronomy, University of 
Pennsylvania, Philadelphia, PA 19104}
\altaffiltext{3}{Alfred P. Sloan Fellow}

\begin{abstract}

We use the redshift evolution of the \lya forest mean transmitted 
flux at $z\gtrsim 2.5$ to
infer the evolution of the intensity of the ionizing background,
using theoretical predictions for the 
density distribution of the intergalactic gas giving rise to the 
\lya forest. The proper background intensity declines gradually with
redshift, decreasing by a factor $\sim 3$ between $z=3$ and $z=5$.

 The gravitational structure formation theory of the \lya forest implies
that, given the observed mean transmitted flux of $\bar F = 0.1$ in the
highest redshift quasar (with a mean \lya forest redshift $z=5.2$),
only about $3\%$ of the \lya spectrum should have a flux higher than
$50\%$ of the continuum, and less than 0.5\% of pixels should have 
flux above $75\%$ of the continuum, assuming a uniform ionizing
background intensity. We show that this is consistent with the spectrum
shown in Fan \etal.

\end{abstract}

\keywords{
cosmology: theory---intergalactic medium---large-scale structure of
universe---quasars: absorption 
lines
}

\section{INTRODUCTION}

  The theory of the \lya\ forest based on structure formation in Cold
Dark Matter models has been highly successful in accounting for the
observable properties of the \lya\ forest, in particular the flux
distribution and the power spectrum (see Croft \etal 1999, McDonald
\etal 2000a, and references therein). As new quasars at ever higher
redshift are discovered, the comparison of these observable quantities
with the theory should be extended. Because the theory predicts the
evolution of the spatial distribution of intergalactic gas, it is now
possible to infer the evolution of the ionizing background intensity
from the observed evolution of the mean flux decrement. Furthermore,
as the epoch of reionization is approached, fluctuations in the
background intensity due to the discreteness of the sources should
increase, affecting the \lya\ forest flux distribution.

  In this paper, we use the observations of three quasars at 
$z \geq 5$
(Songaila \etal 1999, hereafter SHCM; Stern \etal 2000; Fan \etal 2000)
to infer 
the evolution in the hydrogen ionization rate, and to test if the flux
distribution is consistent with theoretical predictions.
In \S 2 we present predictions of the flux distribution at high
redshift, using numerical simulations whose resolution and box size are
shown to be sufficient to have produced convergent results.
In \S 3 we compare the predictions to the observations
to measure the evolution of the ionization rate and search 
for differences in the
predicted and observed flux distribution.

\section{NUMERICAL RESULTS FOR THE FLUX DISTRIBUTION}

  If we ignore the effects of peculiar velocity of the gas and of
thermal broadening, then the optical depth to \lya\ scattering, $\tau$,
at any given point of an observed spectrum is determined by the density
of gas at the corresponding point in space, $\rho_{\rm gas}\equiv
\Delta\, (\Omega_b \rho_{\rm crit})$, and by the gas temperature, which
we assume to follow the relation $T=T_0\, \Delta^{\gamma-1}$ (see Hui \&
Gnedin 1997):
\begin{equation}
\tau = \tau_0 ~
\frac{\left(1+z\right)^6~ \left(\Omega_b h^2\right)^2}
{T_0^{0.7}~ H(z)~ \Gamma_{-12}(z)}~\Delta^{\beta}~.
\label{taueq}
\end{equation}
Here, $\tau_0$ is a constant, $H(z)\simeq H_0~ \Omega_m^{1/2}~
(1+z)^{3/2}$ is the Hubble constant at redshift $z$, $H_0$ is the
present Hubble constant, $h=H_0/(100 \kms/\mpc)$, $\Omega_m$ is the
fraction of the critical density in pressureless matter,
$\Gamma_{-12}(z)$ is the photoionization rate of hydrogen in units of
$10^{-12}\, {\rm s}^{-1}$, and $\beta = 2-0.7~(\gmo)$ (where we use
$\alpha(T) \propto T^{-0.7}$ for the recombination coefficient).
The dependence of the distribution and average of the transmitted flux,
$F=\exp(-\tau)$ (which is the observed quantity), on the cosmological
model can be understood from equation (\ref{taueq}). Thus, as long as
the distribution of $\Delta$ and the $\Delta-T$ relation are
sufficiently well understood, the observed mean transmitted flux of the
\lya\ forest can be used to determine the parameter combination
$(\Omega_b h^2)^2/ H(z)/\Gamma_{-12}(z)$ in equation (\ref{taueq}).

  Knowledge of the distribution of $\Delta$ and $T$ needs to come from
the theory, with the use of hydrodynamic simulations. This theory can
be tested by checking its predictions for the detailed form of the
distribution of $F$ and its power spectrum against the observations,
and this comparison has shown that the Cold Dark Matter model with a
cosmological constant that is supported by a large body of
observational evidence in galaxy clustering and CMB fluctuations 
(Primack 2000, and references therein) 
is in good agreement with the observed \lya forest (Rauch \etal 1997,
Croft \etal 1999, McDonald \etal 2000a, and references therein). At the
same time, the gas temperature over the relevant range of densities
has been measured from the absorption line shapes (Ricotti, Gnedin, \&
Shull 2000; Schaye et al.\ 2000; McDonald et al.\ 2000b), although
the systematic differences between these studies are still substantial.

  We focus in this paper on the flux distribution at high redshift.
This particular property of the \lya\ forest is affected only by the
low-density intergalactic gas, given the large mean flux decrement
(the reason is that high density gas causes absorption only in highly
saturated regions of the spectrum). We first show that this distribution
is predicted without significant uncertainties due to the finite
resolution and box size of the simulations. We use the Hydro-PM (HPM)
code of Gnedin \& Hui (1998), comparing it first to the fully
hydrodynamic simulation in Miralda-Escud\'e et al.\ (1996) to test that
the approximate nature of the Hydro-PM method does not result in any
significant differences in the predicted flux distribution at high
redshift. When we create simulated spectra, we assume a 
$\Delta-T$ relation 
given by $T_0=2\times 10^4$ K
and $\gmo=0$, which is consistent with the observational determinations
Ricotti \etal (1999) and McDonald \etal (2000b) at $z\sim 4$, although
Schaye \etal (2000) find a lower temperature (for the calculation
of the pressure term in the HPM simulations we assume $T=0$ before 
reionization at $z=7$, and then interpolate between $T_0=25000$, 
$\gmo=0.0$ at $z=7$ and $T_0=19000$, $\gmo=0.2$ at $z=3.9$; however,
we find that the results are almost completely insensitive to these 
assumptions, probably
because the pressure does not have much influence in the voids).
The cosmological model is $\Lambda$CDM, with $\Omega_m=0.4$,
$\Omega_b h^2=0.015$ (note, however, that the results we give for 
$\Gamma_{-12}$ are rescaled to $\Omega_b h^2=0.02$), 
$\Omega_\Lambda = 0.6$, $h=0.65$, and $n=0.95$,
where $\Omega_\Lambda$ is the fraction of the critical density
in cosmological constant and $n$ is the usual logarithmic slope of
the primordial power spectrum.
Simulated spectra are constructed in which the transmitted flux is binned 
into 2 \AA$~$ pixels, to match the observed spectrum in Fan \etal (2000)
(we use this pixel size throughout the paper). 
In Figure 1, we show the flux
distribution predicted for the full-hydro simulation as the heavy solid
line.  The light 
({\it long-dashed, dotted, solid, dashed}) lines show
HPM simulations with box size
$L=$(17.8, 8.9, 4.4, 4.4), and a total number of particles
$N=(256^3,~ 128^3,~128^3,~ 256^3)$, respectively.
All the
results are at $z=4$, but using the value of $\Gamma_{-12}$ that
gives a mean flux decrement from the hydrodynamic simulation of
$\bar F = 0.1$, which matches the observations at $z\sim 5$. 
We keep $\Gamma_{-12}$ (rather than $\bar F$) constant for all the
simulations, so that we can see the difference in the predicted
$\bar F$. 
\begin{figure}
\plotone{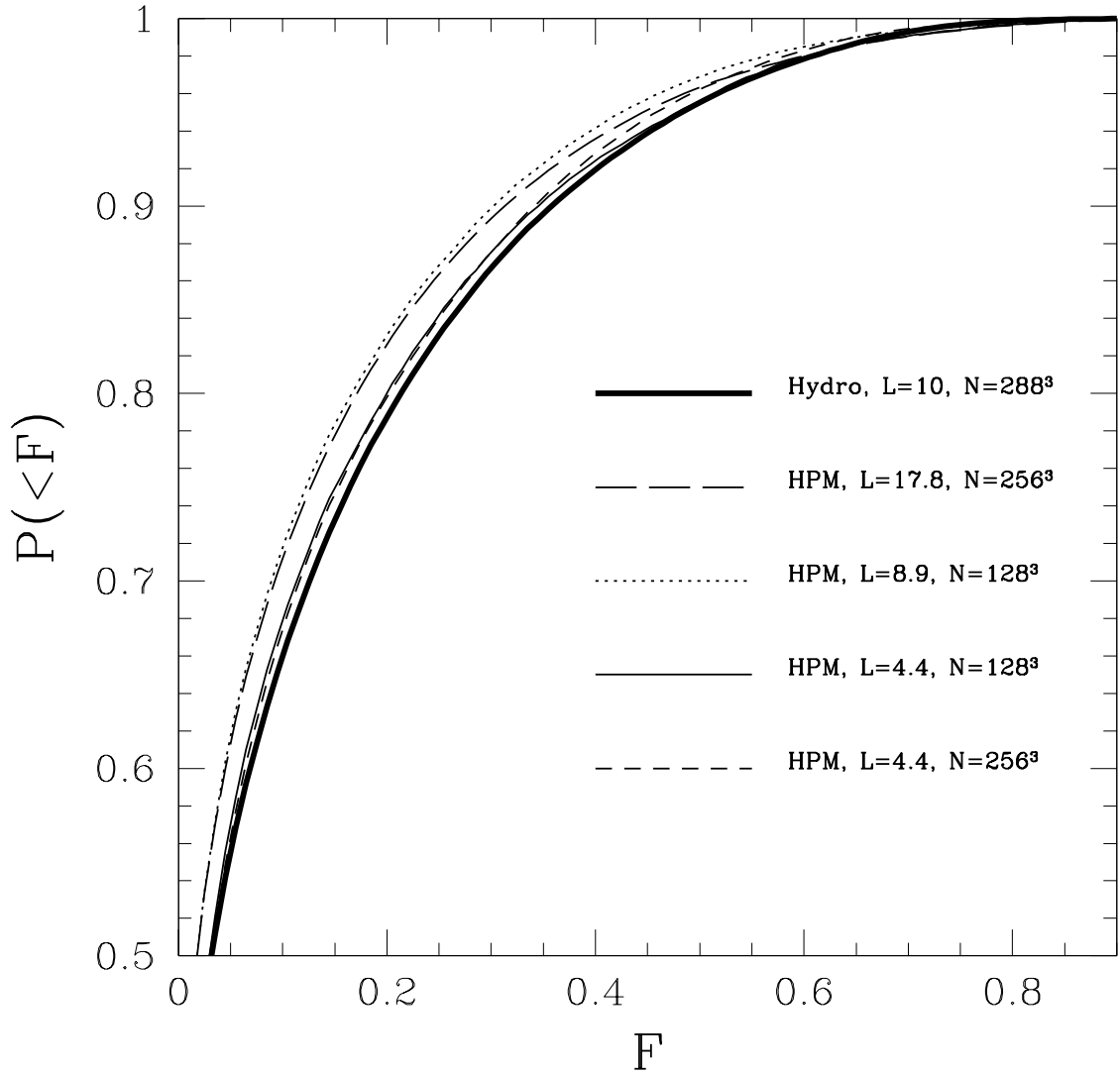}
\caption{Cumulative distribution of transmitted flux.
{\it Heavy solid line}: full hydrodynamic simulation.
{\it Thin lines}: Hydro-PM simulations, with box sizes and particle
numbers indicated.
}
\label{numtests}
\end{figure}

The result for the fully hydrodynamic simulation and for the HPM
simulation that has almost the same box size and resolution
($L=8.9$ and $N=256^3$, not plotted) 
are essentially identical.  This justifies using the HPM
simulations to examine the effects of changing the box size and 
resolution and to compute the flux distribution for comparison with
observations at different redshifts. 
The two simulations with the
smallest box size, with resolutions $L/N^{1/3}=35\hkpc$ and
$17 \hkpc$ (solid and dashed lines, respectively), show that
simulations with inter-particle separation of $35 \hkpc$ have already
achieved the required resolution; however, degrading the resolution
to $L/N^{1/3}=69\hkpc$ causes a change in $\bar F$ of $0.02$, similar
to the observational error bars of $\bar F$.
Comparison of the long-dashed and dotted (or the heavy and light solid)
lines shows that the effect of increasing the box size beyond $5\hmpc$
is small.  We conclude that simulations with boxes $L=8.9 \hmpc$
and number of particles $N=256^3$ are sufficient for our purpose to have
achieved convergence. 
We find similar convergence in the HPM simulations at $z=5.2$.
We also find that different random realizations
of the initial conditions of the simulation of this box size and
resolution change $\bar F$ by $\pm 0.01$ (small compared to the
observational error).

\section{RESULTS}

  We now examine the flux distribution at high redshift predicted from
the simulations to do a direct comparison to the observational
constraints. We use the HPM simulation with $L=8.9\hmpc$ and $N=256^3$, 
and create spectra binned into 2 \AA$~$ pixels. 
\begin{figure}
\plotone{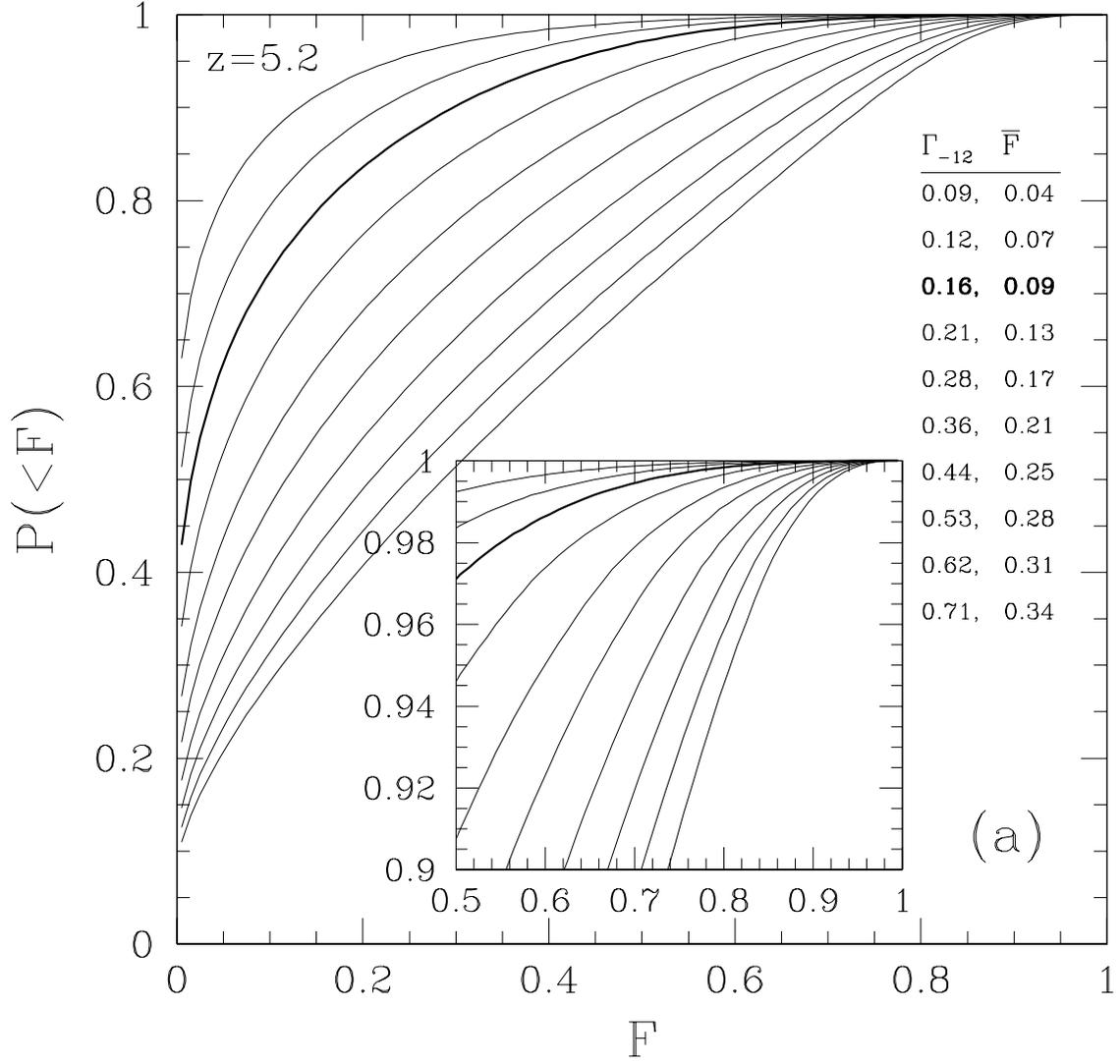}
\caption{Cumulative flux distributions for (a) $z=5.2$ and (b) $z=4.5$.
Listed values of $\Gamma_{-12}$ and $\bar{F}$ apply to the lines from
top to bottom.  The inset box in (a) is an expanded version of the
upper right hand corner of the full figure.
}
\label{highzres}
\end{figure}
\begin{figure}
\plotone{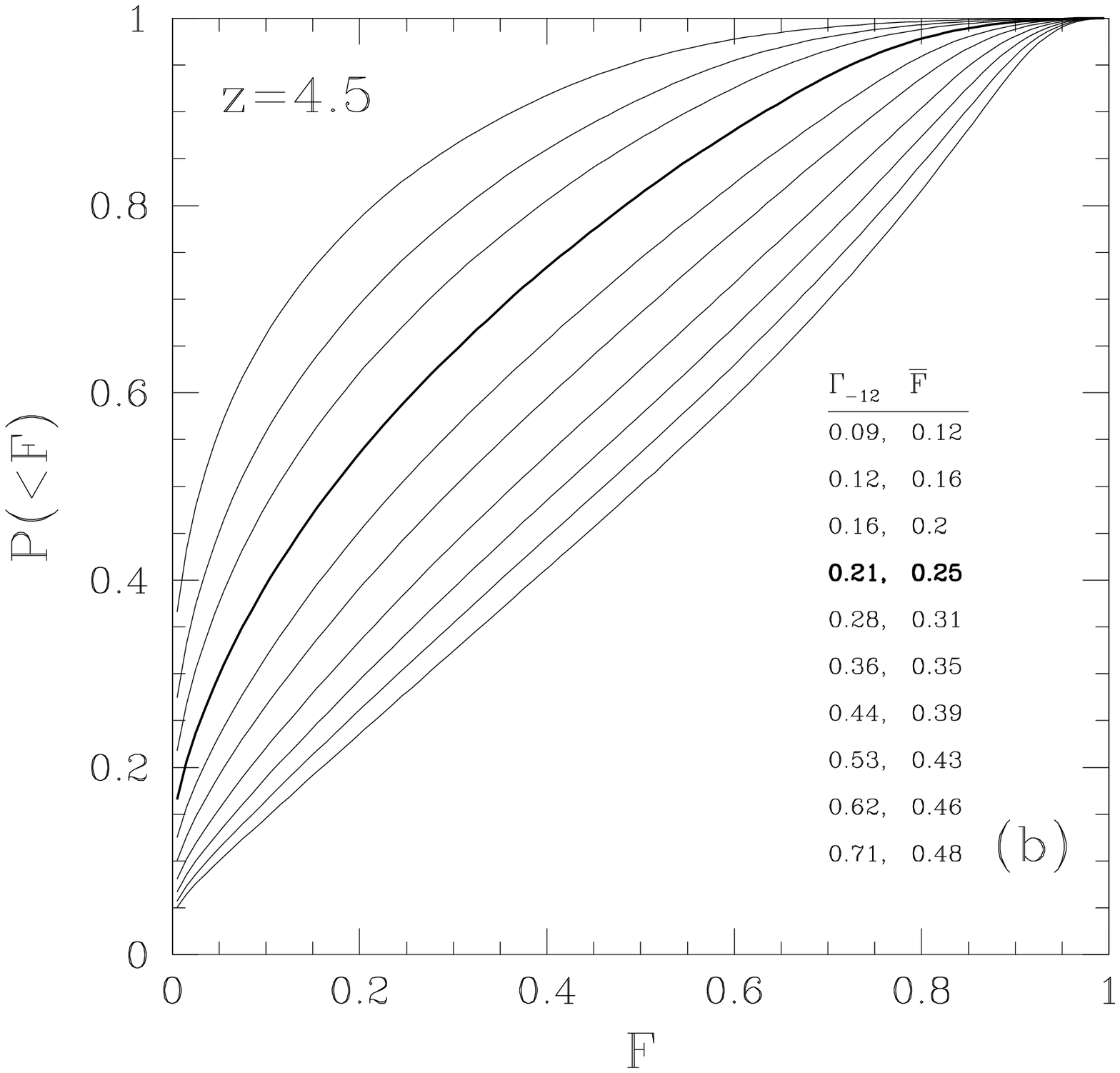}
\end{figure}
The curves in Figure \ref{highzres}(a) show the cumulative flux
distribution at $z=5.2$, for ten values of $\Gamma_{-12}$,
corresponding to different values of $\bar F$, indicated in the Figure.
This can be compared to the spectrum in Figure 3 of Fan \etal (2000),
from a quasar at $z=5.8$ (where the mean redshift of the observed \lya
forest, in the wavelength range 7140-7956 \AA$~$ used 
by Fan \etal 2000 to compute $\bF$, is $5.2$). 
In order to match the observed $\bF=0.10\pm 0.02$,
we need $\Gamma_{-12}\simeq 0.17\pm 0.03$,
for the \lya forest model described in \S 2.

  Remarkably, there should be essentially no pixels in the spectrum at
this high redshift where the flux comes close to the continuum level.
Only 3\% of the spectrum should have a flux higher than 50\% of the
continuum, and less than 0.5\% should have a flux greater than 75\% of
the continuum [see the inset of Fig. \ref{highzres}(a)].
In comparison, Fan \etal (2000) identify the minimum 
absorption in their observed 
spectrum to be $\tau \sim 0.4$ 
(although they estimate that continuum uncertainty and noise 
make this number uncertain by about $\pm 0.1$).
Absorption of $\tau \lesssim 0.4$ or $F\lesssim 0.67$ occurs
in approximately 1\% of the pixels in our simulated spectra, so we
predict about three such pixels to be found in the wavelength 
range 7300-7956 \AA$~$ examined by Fan \etal (2000).
Therefore, the minimum absorption present in this spectrum is consistent
with the predictions of the simple \lya forest model in our simulations.
We find that decreasing $\sigma_8$ by $\sim 15\%$ to match the results
of McDonald \etal (2000a) increases the inferred value of $\Gamma_{-12}$
by $\sim 30\%$, but does not affect significantly our conclusions about
the flux distribution (once the observed $\bar{F}$ is matched).

  Next, the same lines are plotted in Figure \ref{highzres}(b) at
redshift $z=4.5$, which is about the mean redshift of the \lya forest in
the $z=5.0$ quasar spectrum that is shown in Figure 2 of SHCM. The mean
transmitted flux in this spectrum is $\bF=0.25$, which requires
$\Gamma_{-12}\simeq 0.21$ (no error bar was given in SHCM for the value
of $\bar F$). 
According to SHCM, an upper bound on the minimum optical 
depth in their spectrum is $\tau < 0.1$. They 
identify this region as being produced by a void with density
$\Delta \simeq 0.2$, to derive a lower limit to the background intensity
of $\Gamma_{-12} > 2.8$, for $\Omega_b h^2=0.019$.

Our value of $\Gamma_{-12}$, inferred by matching $\bF$,
is much smaller for the following reasons:
Factors of $\sim(20000/5000)^{-0.7}$, $(0.4/1.0)^{1/2}$, 
and $0.65/0.5$ are obtained 
from our different choices of $T_0$, $\Omega_m$, and $h$, respectively
(note:  we assume SHCM used $h=0.5$, but their paper is not
clear on this point).  Our choices place
more conservative lower limits on $\Gamma_{-12}$, and are more
consistent with independent observations.
The remaining factor 2.4 must be traced to the identification by SHCM
of observed $\tau=0.1$ with underdensity $\Delta=0.2$.
In our simulations, $\Delta<0.2$ occurs in 1.7\% of pixels in real 
space, and this probability is increased when the expansion
of voids in redshift space is considered.  Consequently, the minimum
absorption in the 200\AA$~$ region considered by SHCM probably arises
from a void with $\Delta<0.2$.
Furthermore, the peculiar expansion of voids 
reduces the optical depth at fixed density.    
After the inclusion of all these effects through our simulations, 
and using $\Gamma_{-12}=0.21$, we
predict $\sim 0.4$ regions with $\tau<0.1$ in the wavelength 
range $6800-7000~$\AA$~$ identified by SHCM.
Lastly, the minimum absorption statistic is biased toward regions
with strong noise contribution, so the value $\tau<0.1$ is probably not
conservative (our simulations give a completely acceptable
1\% probability for $\tau<0.15$). 

  Finally, the quasar in Stern \etal (2000) at $z=5.5$ has mean 
transmitted flux $\bF=0.10\pm 0.02$. The
ionization rate required to match this $\bF$ is 
$\Gamma_{-12}=0.13\pm 0.03$, at a mean redshift $\bar{z}= 4.93$.
Their spectrum as shown has only $\sim 15$ \AA$~$ resolution so we
do not interpret it further.

\begin{figure}
\plotone{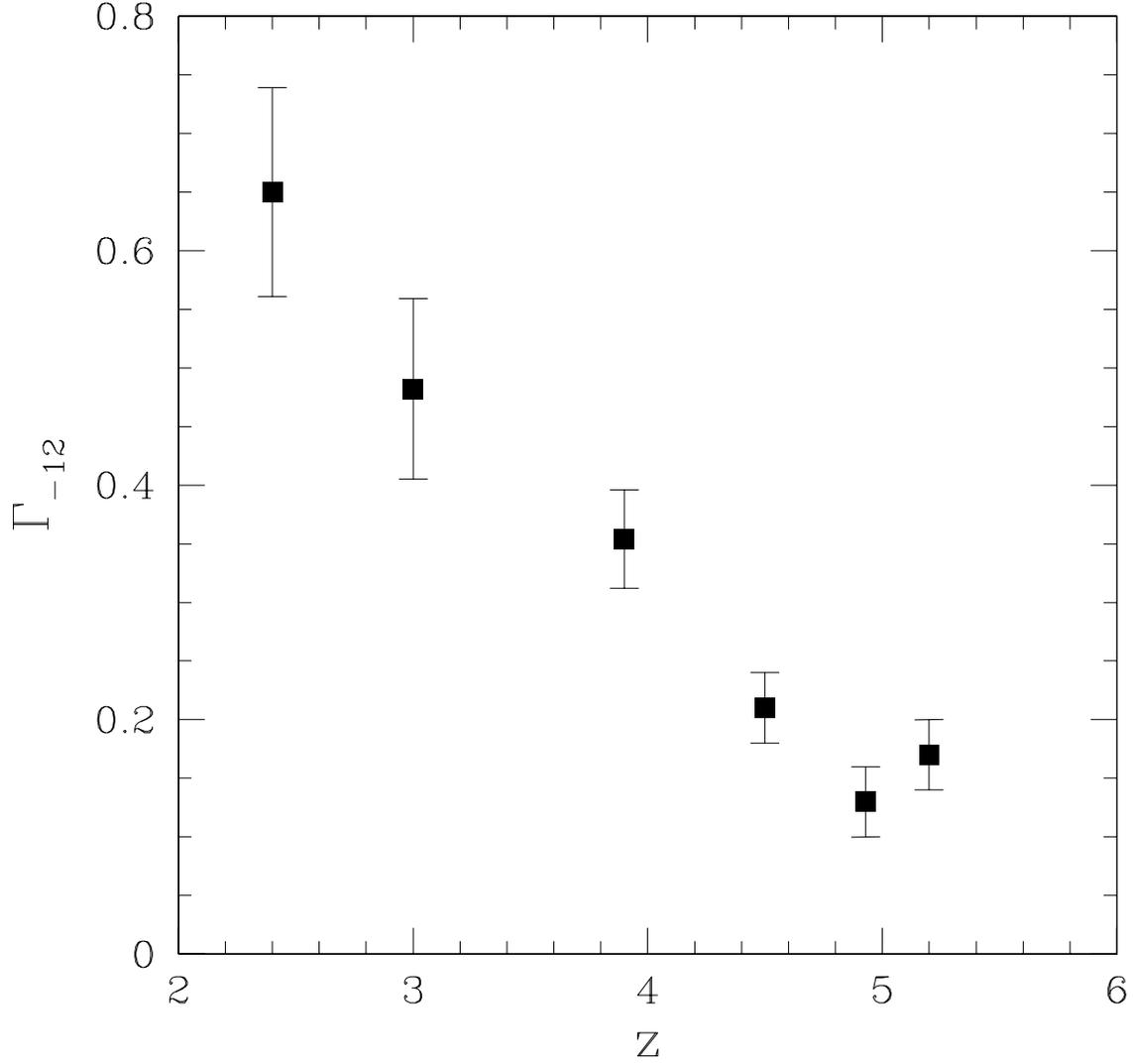}
\caption{Evolution of observed $\Gamma_{-12}$.  
Computed for $\Omega_B h^2=0.02$, and the $\Lambda$CDM model 
discussed in the text (changing $\Omega_B$, for example, will 
shift all the points by the same factor).  
Uncertainties related to the
quasar continuum and the gas temperature may introduce some 
uncertainty beyond the given error bars, 
especially for the 3 highest $z$ points.
}
\label{Gm12results}
\end{figure}

  The evolution of the background intensity $\Gamma_{-12}$ derived
from the observed mean flux decrement, for $\Omega_b h^2 = 0.02$, is
shown in Figure \ref{Gm12results}, using the values from the three
quasars discussed here and the lower redshift results from
McDonald \etal (2000a)
(updated by McDonald \etal 2000b using the measured IGM temperatures).

\section{DISCUSSION}

  There is a clear reduction of the proper intensity of the ionizing
background with redshift, by a factor $\sim 3$ from $z=3$ to $z=5$.
This reduction was noticed previously by Rauch \etal (1997) up to $z=4$.
With the new high redshift quasars, we find a continuation of this
gradual decrease of $\Gamma$, but without a dramatic drop that would
be the sign of approaching the reionization epoch. The consistency of
the fraction of the \lya spectrum with low absorption with the
observations also suggests that the background intensity fluctuations
are not yet extreme at $z=5$.

  The evolution of the background intensity is related to the evolution
of the comoving emissivity, $\epsilon$, and the proper mean free path
$\lambda$, by $\epsilon ~(1+z)^3 \propto \Gamma~ \lambda^{-1}$, if $\lambda$
is small compared to the cosmic horizon. Assuming that the rate of
incidence of Lyman limit systems evolves as $(1+z)^{1.5}$
(Storrie-Lombardi \etal 1994), the mean free path
of ionizing photons is $\lambda \propto (1+z)^{-4}$, and therefore the
comoving emissivity changes between $z=3$ and $z=5$ by a factor
$(1/3)\,(6/4)^{4-3}=0.5$, indicating only a small reduction, which
is clearly different from the reduction by a factor 0.14 predicted by 
the evolution of the comoving density of luminous quasars
(Schmidt, Schneider, \& Gunn 1995)
and implies, therefore, that galaxies or other sources are dominating
the emission of ionizing photons at $z=5$.

We thank Nick Gnedin for 
providing a copy of his Hydro-PM code and for
helpful comments on the manuscript,
Renyue Cen for the output from his hydrodynamic simulation,
and Adam Steed for helpful comments on the manuscript.

\end{document}